\documentclass{article}

\usepackage{arxiv}

\usepackage[utf8]{inputenc} 
\usepackage[T1]{fontenc}    
\usepackage{hyperref}       
\usepackage{url}            
\usepackage{booktabs}       
\usepackage{amsfonts}       
\usepackage{nicefrac}       
\usepackage{microtype}      
\usepackage{lipsum}
\usepackage{graphicx}
\graphicspath{ {./images/} }

\newcommand\tab[1][1cm]{\hspace*{#1}}

\title{Overview of Quantum Key Distribution
Technique Within IPsec Architecture}

\author{
 Emir Dervisevic \\
  Department of Telecommunications, Faculty of Electrical Engineering\\
  University of Sarajevo\\
  Zmaja od Bosne bb, 71000, Sarajevo, Bosnia and Herzegovina\\
  \texttt{emir.dervisevic@etf.unsa.ba} \\
   \And
 Miralem Mehic \\
  Department of Telecommunications, Faculty of Electrical Engineering\\
  University of Sarajevo\\
  Zmaja od Bosne bb, 71000, Sarajevo, Bosnia and Herzegovina \\
}

\begin{document}
\maketitle

\begin{abstract}
Quantum Key Distribution (QKD) is an approach for establishing symmetrical binary keys between distant users in an information-theoretically secure way. In this paper we provide an overview of existing solutions that integrate QKD within the most popular architecture for establishing secure communications in modern IP (Internet Protocol) networks - IPsec (Internet Protocol security). The provided overview can be used to further design the integration of QKD within the IPsec architecture striving for a standardized solution.
\end{abstract}


\section{Introduction}
Achievements in the field of quantum computing are raising concerns about security mechanisms employed today~\cite{arute2019quantum}. Quantum computers should provide higher computing power and some already described quantum algorithms~\cite{yan2013quantum, shor1994algorithms} for such computers, suggest that they will be able to solve many mathematical problems in polynomial time that are now considered to be of exponential time complexity. This would imply that quantum computers would be able to break many known public key cryptosystems~\cite{furrer2020roger}.

Public key cryptosystems are often used to securely establish a shared symmetric keys between distant parties. The examples of such systems are Rivest, Shamir, Adleman (RSA), Diffie–Hellman (DH), Elliptic Curve Cryptography (ECC), and ElGamal. Among them, the DH key exchange denotes the beginning of development for the asymmetric cryptography~\cite{diffie1976new} and is widely used today in a security protocols such as TLS (Transport Layer Security), IPsec (Internet Protocol Security), SSH (Secure Shell), and many others~\cite{furrer2020roger}.

In recent years, notable efforts have been made in development of the quantum safe networks based on Quantum Key Distribution (QKD). QKD, as a name would suggest, is a method for secret key distribution that is based on the laws of quantum physics. It allows distant parties to establish a shared symmetric keys for which secrecy can be guaranteed. As it is not a complete quantum-safe network solution but rather a concept for secret key establishment, QKD's integration with some existing and well-established security architectures is of great importance. IPsec has been used for more then 20 years to establish secure tunnels over public Internet infrastructure. It is  considered as the most popular technique for establishing secure communication in the modern IP networks.  In this paper we provide an overview of solutions that integrate the QKD method within the IPsec architecture. The main goal of this integration is to provide an IPsec engine with the secret keys generated by the QKD. Using QKD keys instead of DH keys to establish QKD secure IPsec tunnels provides a unique way to converge QKD technology with modern IP networks.

The remainder of this paper is organized as follows: In Section~\nameref{sec:qkd} a brief description of the QKD technique is provided. Section~\nameref{sec:ipsec} describes the IPsec architecture and its individual components. An Internet Key Exchange (IKE) protocol, as an important component in the overall IPsec architecture is described in Section~\nameref{sec:ike}. An overview of existing solutions that integrate the QKD technique within the IPsec architecture is given in Section~\nameref{sec:qkd-ipsec}.

\section{Quantum key distribution}
\label{sec:qkd}
Quantum key distribution with quantum safe symmetric encryption, is one of the very few methods that can provide provable security in the post-quantum era. QKD has a potential to provide an information-theoretically secure (ITS) way of establishing secret keys between two distant peers, thus, giving an alternative solution to the secret key agreement problem~\cite{islam2018high, alleaume2014using}. In 1984, Charles Bennet and Giles Brassard described the first QKD scheme~\cite{bennett1984quantum}, now known as BB84 protocol, for secret key distribution by exploiting the laws of quantum physics. Few years later, in 1989, Bennett and Brassard demonstrated the first QKD experiment over a distance of 32.5 cm in freespace~\cite{bennett1989experimental}. This experiment stimulated interest in the integration and wider application of this technology, which is still present today.

\subsection{How does it work?}
The establishment of a secret key using the QKD technique requires special purpose equipment. The two parties involved in the key establishment process require a connection with a QKD link, which is a logical link consisting of a public and a quantum channel. On the quantum channel the secret key is distributed via a stream of independent photons, where each photon carries information about one secret key's bit. Photons, particles of light, show quantum behaviour and those that carry information in their quantum states, such as polarisation, are known as qubits. The quantum behavior of photons allows legitimate parties to detect the presence of an eavesdropper on the quantum channel. This useful property comes from the Heisenberg uncertainty principle. To obtain information carried by the qubit one must perform a measurement, and stated by the uncertainty principle, every measurement perturbs the state of the quantum system - the qubit, thus revealing the presence of the eavesdropper. Moreover, the “no cloning” theorem states that it is not possible to create a perfect copy of the unknown quantum state which further prevents the eavesdropper from obtaining any information without being noticed. However, due to the characteristics of a quantum transfer - the key distribution over the quantum channel, imperfections of the communication medium and the quantum devices, further discussion on the distributed secret key is necessary. This discussion between peers is carried over the public channel, and is know as a post-processing stage in the QKD scheme. In the post-processing stage the parties align the information on the distributed secret key, estimate a quantum error rate, correct the errors, and further strengthen the secrecy of the key. After the post-processing stage, the peers have established the symmetric, shared secret key. As the discussion over the public channel is conveyed in a classical way, it must be authenticated to prevent man-in-the-middle attack.~\cite{kollmitzer2010applied, gisin2002quantum}

\subsection{Limitations}
QKD has several limitations that are worth mentioning. Firstly, although in theory it is the technology that provides the highest degree of security, practical implementation shows noticeable challenges and shortcomings. There is a number of described and demonstrated attacks (such as a Photon Number Splitting (PNS) attack~\cite{huttner1995quantum}, a Trojan Horse attack~\cite{bethune2000autocompensating}, and a Fake states attack~\cite{makarov2005faked}) on QKD that show how limitations of today’s technology can be exploited to successfully obtain information on the secret key without being detected. However, with every presented attack, the engineers come up with a solutions patching exploited loopholes in the QKD system. Another significant limitation of the QKD systems arises from the fact that the key generation rate is interconnected with a distance between two parties involved in the quantum key distribution. A longer distance implies a lower key rate due to imperfections of the communication medium, such as an absorption and scattering of photons in optical fibers~\cite{gisin2002quantum}. The key rates in an advanced QKD systems are only few hundreds of kbps while the length is roughly limited to 150 km~\cite{lucamarini2018overcoming, yuan201810, zhang2020long}, which is significantly low compared to data rates in the current networks. Furthermore, a standalone QKD system is vulnerable to Denial of Service (DoS) attacks. DoS attack can result in a shortage of the QKD keys and force the parties to either stop the secret communication or to use less secure keys that are likely established by public key cryptosystems~\cite{rass2012turning}. 

A limitation on the key rates can be soothed by implementing buffers, storages of generated keys at the both ends of the QKD link. In this manner, QKD can continuously generate keys in advance and store them in buffers from where they can be easily accessed and used~\cite{mehic2017implementation}. Furthermore, a limitation on distance, and vulnerability on DoS attacks of the standalone QKD systems can be overcome by building quantum networks. In such networks, the secret keys between arbitrary, distant peers can be established in a hop-by-hop or key-relay manner~\cite{elliott2007darpa, kollmitzer2010applied}. However, this implies that all network nodes in the QKD network must be trusted. The quantum networks are heavily researched, and a number of testbeds were implemented. For interested readers we recommend~\cite{mehic2020quantum} that provides an overview on a quantum network architectures and testbeds.


\section{Internet Protocol security}
\label{sec:ipsec}
The IPsec is a protocol suite that provides a variety of security services such as integrity, authenticity, confidentiality and more, to the IP traffic. The first standardized specification of the IPsec protocols was published in 1995, in an RFC1825 document called \textit{Security architecture for the internet protocol}~\cite{atkinson1995security}. IPsec operates on a network layer of a TCP/IP (Transport Control Protocol/Internet Protocol) protocol stack~\footnotemark, and thus provides end to end security at all levels of connectivity. Applying security on the network layer has some additional advantages, and probably the most important one is that multiple transport protocols and applications can share a key management infrastructure provided by the network layer~\cite{doraswamy2003ipsec}.

\footnotetext{The TCP/IP protocol stack is the world's most widely-used suite of protocols that are grouped in 4 different layers (application, transport, network and data link layer in that particular order) where each layer has well defined functions and capabilities. Every layer in the stack provides services to the layer above, and uses services offered by the layer below it.
}

\subsection{IPsec architecture}
The IPsec architecture consists of a several databases and protocols. The major databases in the IPsec architecture are: 
\begin{itemize}
    \item Security Policy Database (SPD), and
    \item Security Association Database (SAD).
\end{itemize}

The IPsec protocol suite consist of the following protocols: 
\begin{itemize}
    \item Encapsulating Security Payload (ESP),
    \item Authentication Header (AH), and 
    \item Internet Key Exchange (IKE) protocol.
\end{itemize}

\subsubsection{Security Policy Database}
The SPD database maintains a number of policies that determine which IP traffic flow needs protection, what level of protection should be applied and with whom the protection is shared. Each packet passing through the network layer is first mapped to the IPsec policy based on IP and next layer header information. A policy determines whether the packet should be protected with the IPsec security services, discarded, or bypass the IPsec protection.~\cite{doraswamy2003ipsec}

\subsubsection{Security Association Database}
The SAD is a database in which Security Associations (SAs) are maintained. SAs are an agreement between two parties on how to apply security services to the IP packets. They determine the IPsec protocol used for securing the packets (ESP or/and AH), transforms - bundle of cryptographic algorithms or the protection suite (encryption and/or authentication algorithm for the example), and the secret keys. They are uniquely identified using a Security Parameter Index (SPI) value and the destination address to which they apply. The source can identify the SA using IP and next layer header information, while the destination must resort to the SPI values because it does not have access to such information in a protected packet (this information could be encrypted, and therefore is not available on the destination). Therefore, the SPI value that uniquely identifies the SA on the destination is sent with every packet. To each SA a value of lifetime can be assigned which determines for how long they are valid. This value can be expressed in seconds or kilobytes. If a given time passes, or a given amount of traffic is protected with the SA, it will be replaced by a new one. This process is called rekeying, because a fresh secret key material will be assigned to the new SA that will be used to protect the same underlying traffic as the old one.~\cite{doraswamy2003ipsec}

\subsubsection{Encapsulating Security Payload}
\begin{figure}
  \centering
  \includegraphics[angle=0]{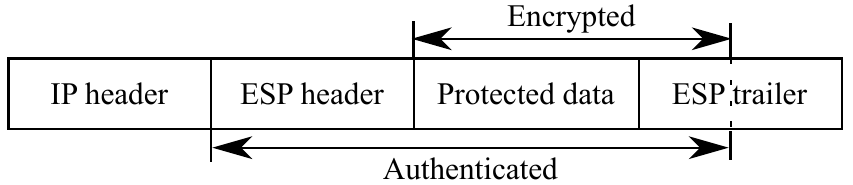}
  \caption{An ESP-protected IP packet}
  \label{fig:esp}
\end{figure}

The Encapsulating Security Payload protocol is in charge of providing confidentiality, data integrity and authentication to the data traffic in the overall IPsec architecture. These tasks are achieved by adding a new protocol header (an ESP header) after an IP header and before data being protected. Moreover, the ESP protocol defines an ESP trailer that is added at the very end of the IP packet. The data between the ESP header and the ESP trailer is protected (see Figure \ref{fig:esp}) and can either be an entire IP datagram or an upper-layer protocol depending on defined IPsec mode\footnotemark.~\cite{doraswamy2003ipsec}

\footnotetext{IPsec defines two modes: a transport and a tunnel mode. The transport mode protects the upper-layer protocols, while the tunnel mode protects the entire IP packets. In case of the tunnel mode being used, the entire protected packet is encapsulated with an outer IP header and sent trough the public network.}

The ESP protocol is a generic and extensible security mechanism. It relies on the construct of the SA which describes the actual cryptographic algorithms and the secret keys for the encryption and/or authentication. Each ESP protected packet will in its header have the SPI value that the destination can use to find the corresponding SA to properly process such packet. In this manner, it is possible to use a newly invented algorithms for the encryption and the authentication and define them in the SA without introducing any changes to the ESP protocol.~\cite{doraswamy2003ipsec}

\subsubsection{Authentication Header}
\begin{figure}
  \centering
  \includegraphics[angle=0]{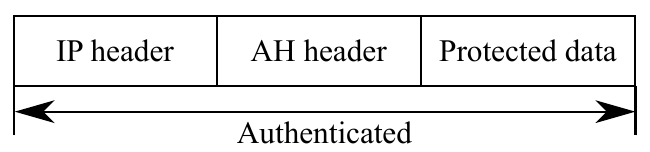}
  \caption{An AH-protected IP packet}
  \label{fig:ah}
\end{figure}

The Authentication Header protocol provides data integrity, authentication, and protection against replay attacks to the IP traffic, however it does not provide confidentiality. Like the ESP, the AH protocol provides mentioned security services to the IP packets by adding an AH header after the IP header and before the data being protected (see Figure \ref{fig:ah}).

The authentication coverage of the AH differs from that of the ESP. The AH authenticates portions of the outer IP header, or more precisely it authenticates those fields of the outer IP header that are not mutable in addition to the authentication of the data being protected.~\cite{doraswamy2003ipsec}

\subsubsection{Internet Key Exchange protocol}
IPsec uses an Internet Key Exchange protocol as a default protocol for the establishment and the management of the SAs. IKE was originally described in RFC2409~\cite{harkins1998internet} as a hybrid of the Oakley and SKEME protocols that is based on an Internet Security Association and Key Management Protocol (ISAKMP) framework originally described in RFC2408~\cite{maughan1998internet}. It is a general-purpose security exchange protocol and thus can negotiate parameters and exchange the secret keys for a different services, including IPsec. More details about the usage of IKE in IPsec can be found in RFC2407~\cite{piper1998rfc2407}.

To exchange the secret keys, IKE uses the DH key exchange method. Therefore, as one of the purposes of the IKE protocol (exchange of the secret keys) overlaps with the main purpose of QKD, one can come to the conclusion that the IKE protocol is the main component of interest in the overall IPsec architecture that should be replaced or modified to integrate QKD with IPsec.


\section{The Internet Key Exchange protocol}
\label{sec:ike}
The main purpose of the IKE protocol in the overall IPsec architecture is to negotiate the IPsec SAs. In general, the establishment of the IPsec SAs in the IKE protocol is performed in two phases, simply referred to as a phase 1 and a phase 2, and can be described as follows:

\begin{itemize}
    \item \textbf{The phase 1.} In the phase 1, an IKE SA (often referred to as an ISAKMP SA) is negotiated between endpoints and the secret keys are exchanged. Furthermore, a mutual authentication of the identities and the exchanged secret keys is performed. The construct of the IKE SA is similar to the IPsec SA, however one must not confuse one with the other. The IPsec SA protects the actual data on the network layer, while the IKE SA protects the IPsec SA negotiations. Therefore, as a result of the phase 1, a secure control channel between two endpoints is established. 
    \item \textbf{The phase 2.} Under the protection of the IKE SA, the negotiation of the IPsec SAs is performed in the phase 2. The secret keys for the negotiated IPsec SAs are derived from the IKE SA secret key if a Perfect Forward Security (PFS)\footnotemark is not requested. Otherwise, a new exchange of the secret keys will take place in the phase 2. The negotiated IPsec SAs protect the IP traffic for a certain time that is defined by the value of their lifetimes. When the IPsec SA expires, a new one is negotiated under the protection of the same IKE SA.
\end{itemize}

\footnotetext{The PFS is defined in RFC 2409~\cite{harkins1998internet} as a restriction of deriving any additional keys from the previously used key.}

\subsection{IKE version 1}
The IKE version 1 (IKEv1) defines a main and an aggressive mode exchange for the phase 1 negotiation and a quick mode exchange for the phase 2 negotiation. The main mode can provide a better security while the aggressive mode can be completed much quicker. The IKEv1 protocol defines two additional exchanges: an informational exchange in which endpoints can communicate error and status information, and a new group exchange that allows endpoints to negotiate the use of custom DH groups~\cite{doraswamy2003ipsec}. Here we describe the main and the quick mode exchange in more detail.

\subsubsection{Main mode exchange}
In the main mode exchange there are a total of six ISAKMP messages exchanged between peers to complete the phase 1. The ISAKMP messages are constructed by chaining multiple ISAKMP payloads together to an ISAKMP header. The ISAKMP header and the payloads are described by an ISAKMP framework and their format will not be discussed in this paper. Important features of the main mode exchange are identity protection and the ability to negotiate DH groups, at least those that are specified by the IKE protocol. IKE allows different authentication methods to be employed, thus the main mode exchange will slightly differ based on the agreed method. The acceptable methods of the authentication are: preshared keys, digital signatures using the Digital Signature Algorithm (DSA), digital signatures using the Rivest-Shamir-Adelman (RSA) algorithm, and two similar methods of authenticating via exchange of encrypted nonces.~\cite{doraswamy2003ipsec}

\begin{figure}
  \centering
  \includegraphics[angle=0]{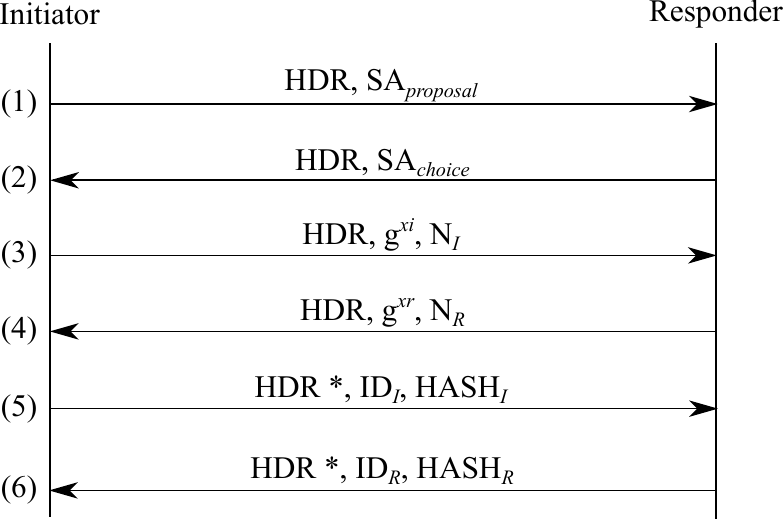}
  \caption{Main mode exchange with preshared authentication}
  \label{fig:main-mode}
\end{figure}

In this paper, the main mode exchange with the preshared authentication method being employed is described, as it has the greatest potential to be used with QKD. The sequence diagram of the main mode exchange with preshared authentication is shown in Figure \ref{fig:main-mode}. The ISAKMP header is denoted with HDR, and a symbol * that follows HDR is an indication that the payloads that follows are encrypted. The main mode exchange shown in the Figure \ref{fig:main-mode} can be described as follows:

\textbf{(1)(2):} The initiator and the responder exchange cookies, CKY-I and CKY-R respectively, via corresponding fields of the ISAKMP header. These values are later used to uniquely identify the established IKE SA. Furthermore, the IKE SA negotiation takes place. The initiator can propose a complex SA offers and the responder chooses one of the multiple proposals.

\textbf{(3)(4):} The initiator and the responder exchange the DH public values, g$^{xi}$ and g$^{xr}$ respectively, and a pseudo-random numbers, N$_I$ and N$_R$ respectively. Based on these values both peers form a secret, called SKEYID. The calculation of the SKEYID is dependent on the authentication method being employed, and for the preshared authentication is given as follows: 

\tab SKEYID = PRF (preshared-key, N$_I$ | N$_R$)

where PRF is a pseudo-random function, usually the HMAC version of the negotiated hash function, and | denotes concatenation. Three keys are further derived from the SKEYID: SKEYID\_d - from which the session keys for the IPsec SAs are derived, SKEYID\_e - used for the encryption of the messages on the control channel, and SKEYID\_a - used for the authentication of the messages on the control channel. 

\textbf{(5)(6):} The initiator and the responder exchange their identities, ID$_I$ and ID$_R$ respectively. Furthermore, the authentication of the main mode exchange as a whole is performed using a keyed hash values. As the payloads are encrypted, the identities of the peers are protected.~\cite{doraswamy2003ipsec} 

\subsubsection{Quick mode exchange}
\begin{figure}
  \centering
  \includegraphics[angle=0]{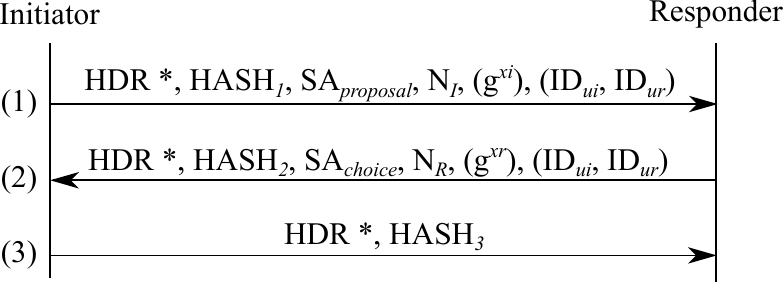}
  \caption{Quick mode exchange}
  \label{fig:quick-mode}
\end{figure}

The negotiation of the IPsec SAs is performed in the phase 2 using the quick mode exchange. The sequence diagram of the quick mode exchange is shown in Figure \ref{fig:quick-mode}, where the brackets indicate an optional payloads, and ID$_{ui}$ and ID$_{ur}$ indicate a proxy negotiation.

The payloads are encrypted using the SKEYID\_e and an encryption algorithm negotiated for the IKE SA. Furthermore, in the quick mode exchange, a hash payload always follows the ISAKMP header, and authenticates the message. The authentication is achieved using the SKEYID\_a and a hash algorithm negotiated for the IKE SA. If the PFS is not requested, the peers will not exchange the DH public values and the session keys for the negotiated IPsec SA will be generated using the SKEYID\_d as the main source of the entropy. The session key of the negotiated IPsec SA is usually denoted as KEYMAT and is calculated as follows (the variables in the square brackets are optional):

\tab KEYMAT = PRF (SKEYID\_d, [ g$^{xy}$| ] protocol | SPI | N$_I$ | N$_R$)

where the SPI values are exchanged during the SA negotiation, a protocol is a value of the IPsec protocol being negotiated, and g$^{xy}$ is the secret key derived from the DH key exchange. The SPI and the protocol value make the IPsec SA unidirectional and protocol specific. The quick mode exchange can be considered as not only a way to negotiate the IPsec SAs, but also as a rekeying mechanism. However, rekeying is technically a new IPsec SA negotiation that is performed when an already existing IPsec SA expires.~\cite{doraswamy2003ipsec, cheng2001architecture}

\subsection{IKE version 2}
The IKE version 2 (IKEv2), introduced in 2005, brought some major changes to the operation of the IKEv1 protocol that made them incompatible. IKEv2 defines four exchanges : IKE\_SA\_INIT, IKE\_AUTH, CREATE\_CHILD, and INFORMATIONAL. In IKEv2, the IPsec SAs are referred to as Child SAs. The concept of the two phase operation still holds, however in IKEv2 there is no clear line as in IKEv1. 

For the establishment of the IKE SA, both the IKE\_SA\_INIT and the IKE\_AUTH must be performed, and in that particular order. In the IKE\_SA\_INIT exchange, the IKE SA negotiation takes place, the DH public values and the pseudo-random numbers are exchanged, and a shared secret SKEYSEED is generated. A seven other secrets are further derived from the SKEYSEED: SK\_d - from which the session keys for the Child SAs are derived, SK\_ai and SK\_ar - used to authenticate messages on the established control channel, SK\_ei and SK\_er - used to encrypt/decrypt messages on the control channel, and SK\_pi and SK\_pr which are used during the IKE\_AUTH exchange. In the IKE\_AUTH exchange, peers exchange their identities and perform a mutual authentication. Furthermore, they negotiate the first Child SA. Therefore, in particular cases, a total of four messages, that is two exchanges, result in the establishment of the IKE SA and the first Child SA. Additional Child SAs can be established using the CREATE\_CHILD exchange. Furthermore, the CREATE\_CHILD exchange is used to rekey both the IKE SA and the Child SA.~\cite{kaufman2010internet}


\section{Existing solutions that integrate QKD with IPsec}
\label{sec:qkd-ipsec}
There are many known approaches to the IPsec implementations that employ QKD derived keys. Several selected approaches that are based on modification of IKEv1 and IKEv2 are discussed. 

\subsection{DARPA quantum network}
\label{sec:darpa}
\begin{figure}
  \centering
  \includegraphics[width=0.88\textwidth, angle=0]{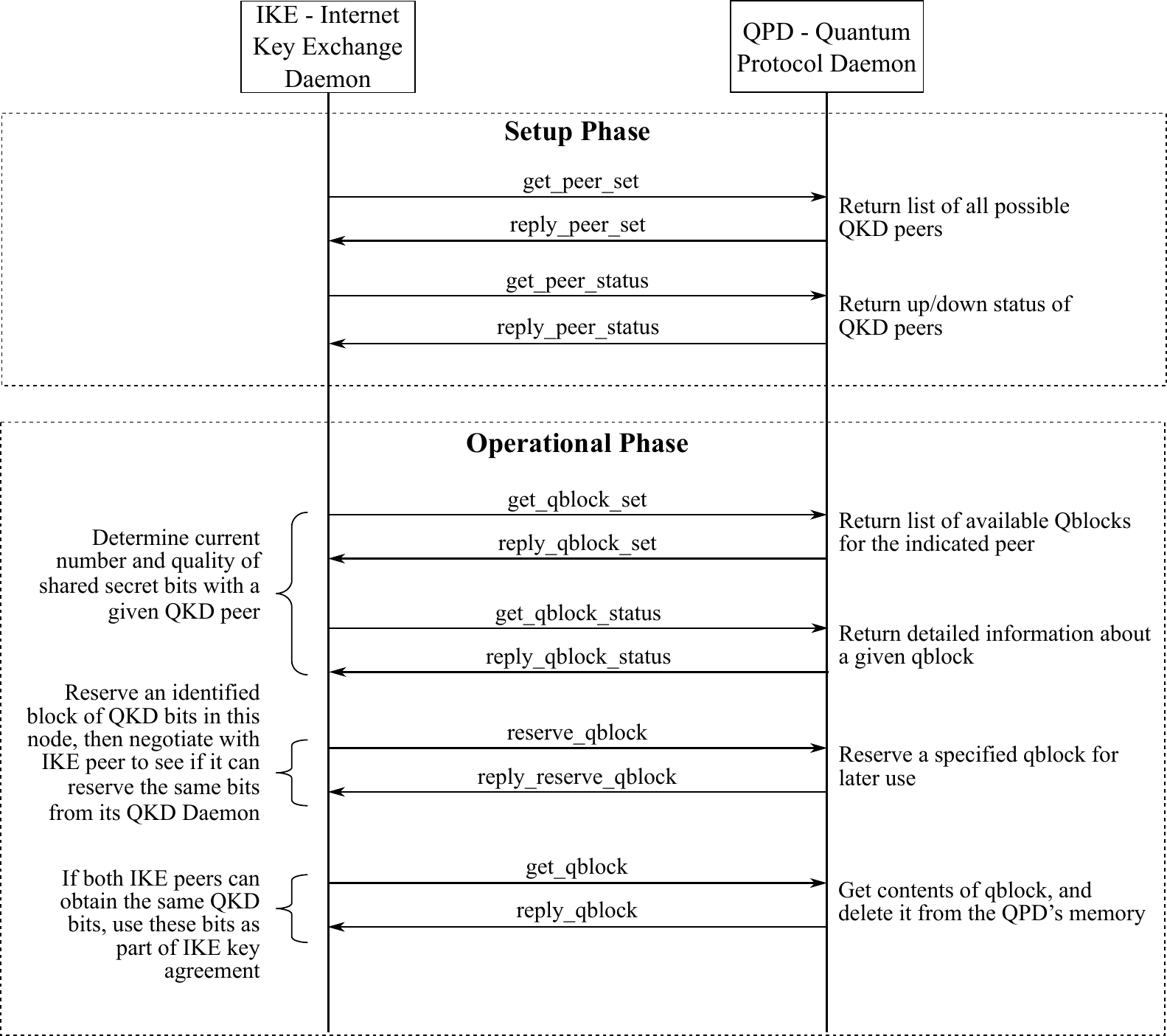}
  \caption{The overview of the IKE / QKD interface in DARPA quantum network}
  \label{fig:ike-qkd-interface}
\end{figure}

In~\cite{elliott2003quantum, elliott2007darpa}, extensions to IKEv1 and IPsec are briefly described to support the use of QKD derived keys in the world's first quantum network - DARPA quantum network. Modifying a kernel and a ‘racoon’ IKE daemon, the following QKD extensions were introduced: a rapid-reseeding and an one-time pad extension. In the rapid-reseeding extension a simple modification in the IKE phase 2 was presented. Distilled QKD bits are negotiated and included in the hash function from which the session keys - KEYMATs are derived. The session keys are then employed as seeds for conventional symmetric ciphers and are updated about once a minute. The use of DH or QKD is completely independent and one can employ the both techniques to generate KEYMAT. In the one-time pad extension, IKE and IPsec were modified to support the use of the one-time pad encryption using a sequence of QKD bits. To serve the IKE daemon with the QKD keys, an IKE / QKD interface was developed. This interface allowed the IKE daemon not only to reserve and obtain the secret QKD keys, but also to obtain a relevant information about the states of the QKD links. The overview of the IKE / QKD interface is shown in Figure~\ref{fig:ike-qkd-interface}, where a Qblock is a fixed-size block of shared secret bits. It is important to note that no modifications were introduced to the IKE phase 1 at the time. It is stated that the level of security offered by QKD is not required for the IKE SA as it only provides protection to non-confidential control messages.

\subsection{MagiQ Technologies}
\label{sec:megiq}
\begin{figure}
  \centering
  \includegraphics[width=0.65\textwidth, angle=0]{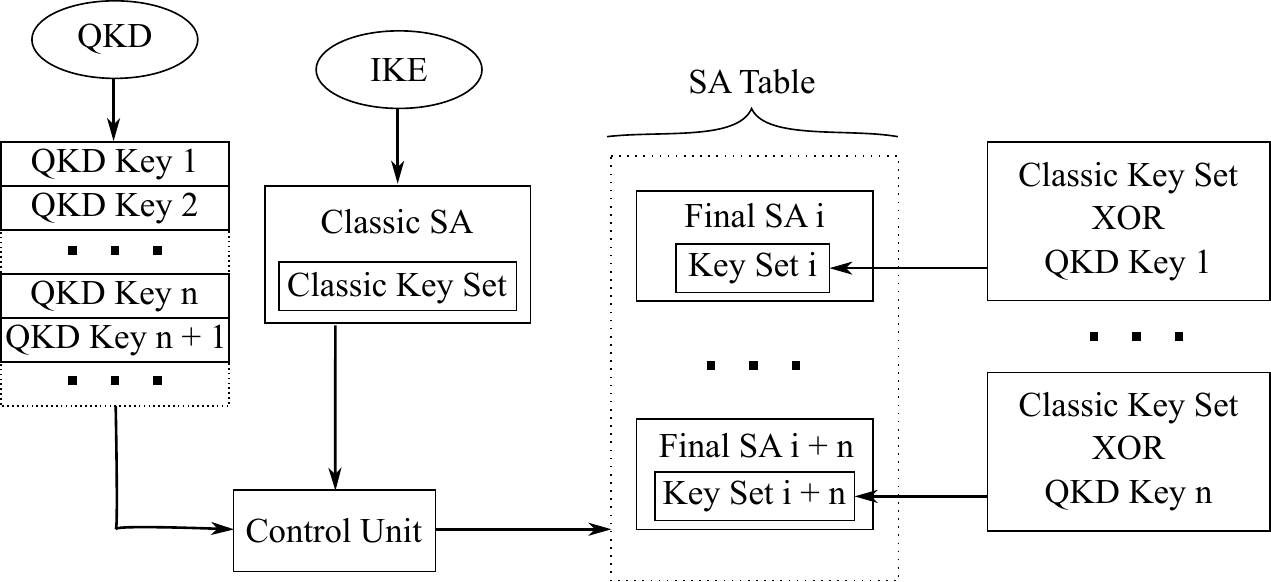}
  \caption{A schematic diagram illustrating establishment of the multiple (n) SAs combining the QKD keys with the classical key generated by the IKE protocol (MagiQ Technologies solution)}
  \label{fig:us-patent}
\end{figure}

MagiQ Technologies, a research and development company, holds a patent~\cite{berzanskis2009method}, solution to achieve a higher frequency at which keys can be changed in the IPsec implementation to improve the overall security. In the standard implementation of IPsec, the two SA tables are created for the particular IP flow, one for the outbound and one for the inbound direction. However, each SA table contains a maximum of two SAs at a time. Extending capacity of each SA table to up to a maximal 2$^{16}$ SAs, this solution allows the fast rekeying. The presented solution combines (e.g., XOR) QKD derived keys with the classical keys generated by the IKE protocol creating multiple SAs in each SA table (see Figure~\ref{fig:us-patent}). This is a local operation and therefore multiple SAs can be created in a very short time. The resulting keys have a greater or equal entropy compared to the classical keys, thus, providing stronger security. Furthermore, a large number of the SAs in the SA tables ensures that the SAs are not being removed before the all reasonably delayed packets are decrypted. 

\subsection{Secure Quantum Key Exchange Internet Protocol}
\label{sec:seqkeip}
In~\cite{sfaxi2005using}, an ISAKMP based protocol is proposed by the name of SeQKEIP (Secure Quantum Key Exchange Internet Protocol). It couples quantum derived keys with IPsec to secure MAN (Metropolitan Area Network) communications. The SeQKEIP runs similar to the IKE protocol but it has additional phase - a phase 0, and a modified IKE phase 1 and IKE phase 2. The phase 1 of SeQKEIP defines a main mode exchange, while the phase 2 defines a quick mode exchange, similar to the IKEv1 protocol. For the phase 0 a new exchange mode, a quantum mode, is presented. In the phase 0, the secret key is exchanged between the nodes using the QKD technique. This key is used as a pre-shared secret in the IKE mechanism. The SeQKEIP uses a pre-shared secret method of the authentication during the phase 1, and it does not define DH groups, nor does it exchange any DH values during this phase. The exchanged key in the phase 0 is used for encryption and authentication of the messages. To support the quantum cryptography concept to full extent, the one-time pad encryption is supported. 

\subsection{AQUA}
\label{sec:aqua}
Under the Advancement in Quantum Architecture (AQUA) project, Nagayama and Van Meter produced an Internet Draft document~\cite{nagayama2009internet}, a solution to integrate QKD with IPsec based on a modified IKEv2 protocol. In 2014. they produced an update to the original draft~\cite{nagayama2014internet}. The proposed solution defines two new payloads in the IKEv2 exchanges: a QKD KeyID payload and a QKD Fallback payload. Furthermore, a new transform field in the SA payload is defined to indicate the use of QKD derived keys. The QKD KeyID payload exchanges a unique identifier (KeyID) of the secret QKD key and an identifier of the QKD device that generated the corresponding key. The secret key is uniquely identified by those two values. If there are no available keys generated by the quantum key distribution protocols, the negotiation in IKEv2 resorts to a fallback method defined in the QKD Fallback payload. The following fallback methods are defined: WAIT\_QKD, CONTINUE, and DIFFIE-HELLMAN. The WAIT\_QKD indicates that IKE must wait for the QKD protocol to generate a new key. The CONTINUE indicates that the most recent key may be used until a new key becomes available, and the DIFFIE-HELLMAN method indicates that IKE shall generate a new key using DH key exchange. It is important to note that the fallback feature is disabled during the IKE\_SA\_INIT and IKE must wait for QKD to generate keys in this case. The IKE\_SA\_INIT exchange, in case when the QKD level of security is desired, includes the KeyID payload and omits the key exchange (payload that carries DH values) and nonce payloads from the standard IKE\_SA\_INIT exchange. Furthermore, it is important to include the new transform payload in the SA proposal. The fallback method is negotiated during the IKE\_AUTH exchange with the use of the QKD Fallback payload. 

\subsection{QIKE}
\label{sec:qike}
\begin{figure}
  \centering
  \includegraphics[width=0.88\textwidth, angle=0]{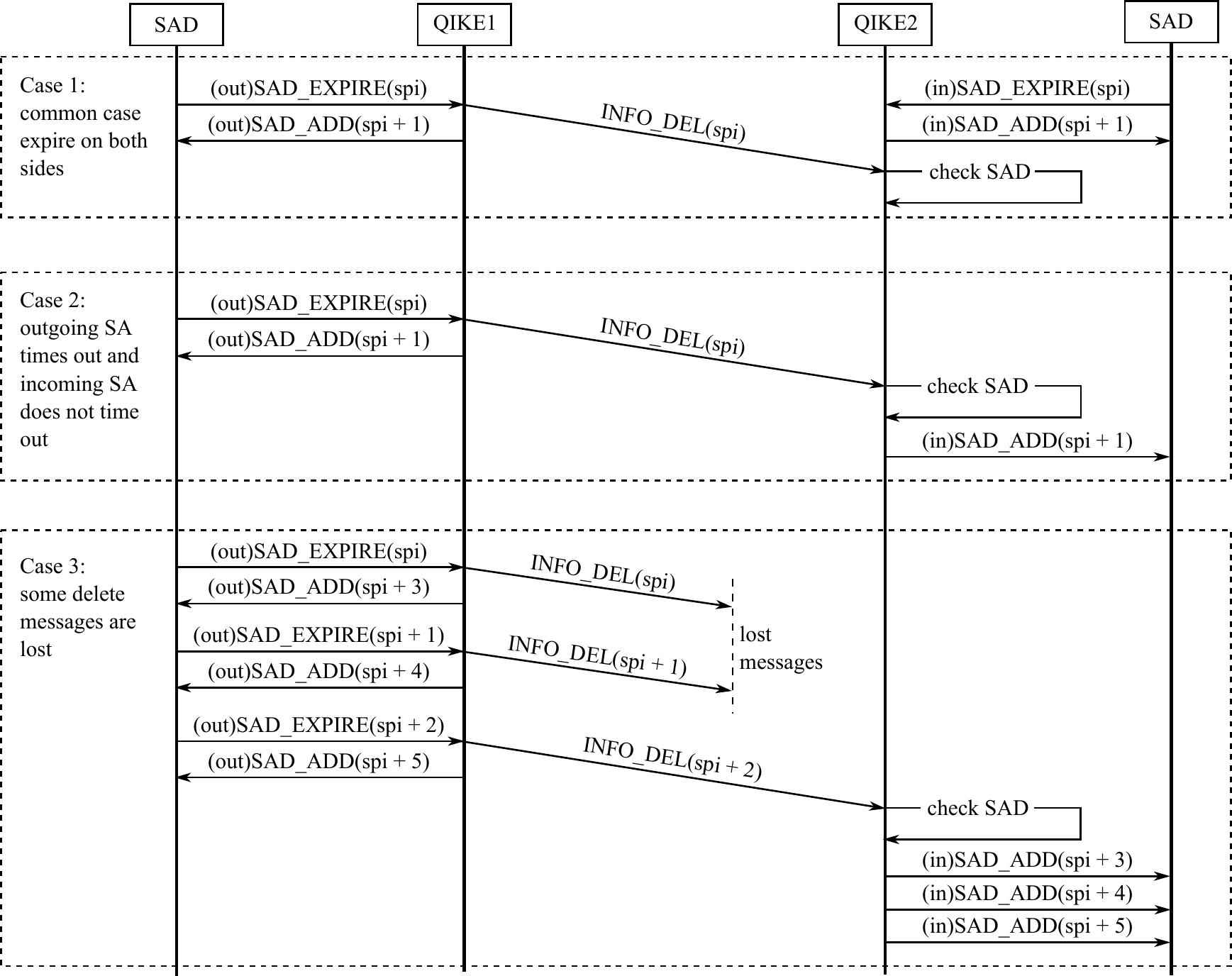}
  \caption{Delete messages and SAD synchronization in QIKE protocol}
  \label{fig:qike-rekey}
\end{figure}

In~\cite{neppach2008key}, a key management solution for the QKD networks is proposed with a primary focus on the fast rekeying in the implementation of IPsec. In the proposed design for the key manager, an application first registers submitting an application ID and then sends connection request which includes the maximum key length that will be requested and the desired key rate. If the requested key rate can be satisfied, the key manager establishes a dedicated key buffer, filled with the QKD key material, for the application. After the connection phase, the application can request the secret QKD keys of desired length, or the information about the state of the key buffer. Further, this paper propose an adaptation of the IKEv1 protocol, called QIKE, to negotiate the SA parameters and requirements on the quantum generated keys. From the key manager point of view, the QIKE protocol is an application that consumes the secret QKD keys. The connection to the key manager is triggered by the QIKE protocol during the phase 1. Therefore, in addition to the standard parameters being negotiated, a QKD key rate and an application ID parameter negotiation in the phase 1 is introduced to establish the IKE SA. The phase 1 messages are authenticated using the configured shared secret (the pre-shared method of authentication).  In QIKE phase 2, similar to the phase 1, the QKD key rate and the application ID parameter are negotiated to establish the IPsec SA. Additionally, a number of the maximum SAs supported by the SAD is negotiated to enable the high speed IPsec implementation. The phase 2 results in a separated buffers for each direction that are allocated at the key manager. The value of lifetime assigned to the each SA in the SAD is given in bytes to solve the synchronization problems of the lifetime defined in milliseconds. If the lifetime of the SA expires, the QIKE protocol does not perform renegotiation of the expired SA. Instead, QIKE uses the ISAKMP informational messages that carry delete payloads to inform a remote QIKE daemon of the expired SA (see Figure~\ref{fig:qike-rekey}). The expired SA is replaced with a new SA that will have the same parameters as the old one, but with a new SPI identificator and a newly assigned QKD key. To deal with the lost of the informational messages, the SPI value is increased per connection to allow a resynchronization of keys in the buffer.

\subsection{Rapid rekeying protocol}
\label{sec:rrp}
\begin{figure}
  \centering
  \includegraphics[width=0.65\textwidth, angle=0]{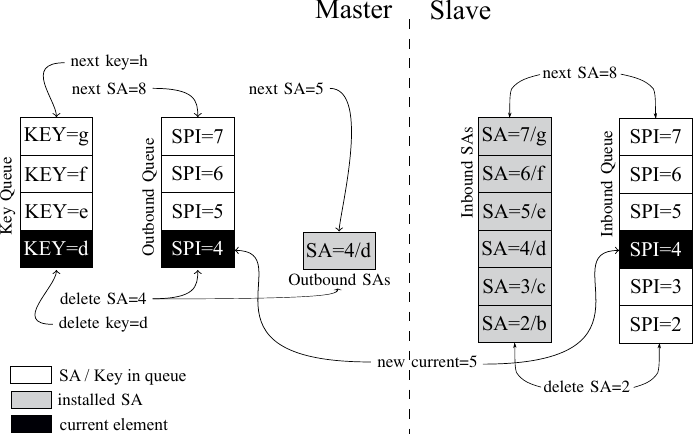}
  \caption{Rekeying procedure in Rapid rekeying protocol}
  \label{fig:rrp-key-change}
\end{figure}

Different approach to implement QKD with IPsec was presented in~\cite{marksteiner2014approach, marksteiner2015protocol} by introducing a specialized, lightweight key synchronization protocol - a Rapid rekeying protocol. The main idea was to refrain from the IKE protocol as there is no need for the classical key exchange to take place. The goal of the designed protocol is to achieve the fast rekeying with purely quantum derived keys in the IPsec implementation. Described protocol uses two channels: an AH - authenticated control channel and an ESP - encrypted data channel. On the control channel a non-secret information is exchanged to ensure a synchronization of the QKD keys during the key change procedure. While the non-secret information on the control channel does not require confidentiality, the authenticity is crucial for the security, and therefore the AH protocol is employed. The rapid rekeying protocol follows master/slave paradigm, where every peer assumes the master role for its outbound connections. The master maintains two queues, one for the precalculated SPI values, and the other for the QKD keys. When the key change procedure takes place, a new pair of the QKD key and the SPI value is added to the queues, and the oldest pair in the queues is assigned to a new active SA. All the other parameters of an expired SA are inherited to the new one. The expired pair is removed from the queues, and the key change request is sent to the peer. Every key change request contains a new SPI value, thus, the receiver can install the SA beforehand. Moreover, a number of SAs exist simultaneously on the receiver side, allowing it to process packets protected with both an older or newer SA then the current one. The oldest SA is deleted on the receiver side, and acknowledgement is sent as a response on the key change request. The acknowledgement serves as keepalive mechanism, as the receiver can compensate the lost of the key change request, and the sender can continue with secure communication without receiving response. The key change procedure is shown in Figure~\ref{fig:rrp-key-change}.  

\section{Discussion and conclusion}
IPSec is the most popular solution for secure communication over a public network. It consists of several subprotocols that are in charge of negotiation and establishment of a secure channel for the IP traffic flow. The main drawback of the IPsec protocol is a dependence on asymmetric cryptographic solutions that are at risk with the upcoming development of quantum computers. In this paper, we provided a detailed overview of the existing solutions that enhance the security offered by IPsec with QKD which offers an approach to the realization of secure communication in the post-quantum era.

The \nameref{sec:darpa} solution provided an IKE / QKD interface, and an ability to derive the session keys from the QKD keys. However, this solution did not address the slow rekeying rates. Moreover, the security provided on the control channel relies on the classical cryptography. The \nameref{sec:seqkeip} utilize the existing method of the preshared authentication supported with the QKD keys providing security even on the control channel. The session keys can be changed at the rate at which the keys are generated by the QKD. However, this approach significantly impacts data rates if the one-time pad encryption is employed as the QKD keys are exchanged on demand (all other solutions assume the QKD keys are exchanged in advance and stored at the endpoints). The \nameref{sec:megiq} solution allows the fast key change rates by allowing a large number of the SAs in the SA tables. Multiple SAs can be established in a very short time, combining the classical keys with the QKD keys. Similar, the \nameref{sec:qike} protocol allows multiple SAs to achive the fast rekeying. The operation of the QIKE protocol is supported with the key manager solution that allows reservation of the QKD keys by the applications. QIKE relies on the ISAKMP informational messages to synchronize the SAD databases, where it is sufficient to exchange the SPI values of the expired SAs. The \nameref{sec:aqua} solution provides extension to the IKEv2 protocol to negotiate the QKD keys. Moreover, this solution allows negotiation of the fallback methods. However, as in the \nameref{sec:darpa} solution, the issue of the slow rekeying is not addressed. The \nameref{sec:rrp} takes a different approach to couple QKD with IPsec. It is a lightweight key synchronization protocol, with the main purpose to keep the QKD keys at the both endpoints synchronized. However, the parameters of the SAs must be known in advance at the both endpoints, as the protocol does not support the negotiation itself. 

The presented solutions show that the use of the QKD keys is achievable in the IPsec architecture. Thus, IPsec could maintain its popularity in protecting critical infrastructures in the post-quantum era. However, to achieve the high rekeying rates, the IPsec architecture should allow a number of the SAs in the SA tables and a mechanism to keep the QKD keys synchronized at both endpoints. 

The main contribution of this paper is providing a summary overview of techniques used for the integration of QKD with common network protocols such as IPsec. Our further work will include practical experiments to convergence QKD with common network applications and protocols used in everyday life.

\section{Acknowledgment}

The research leading to the published results was supported by the NATO SPS G5894 project QUANTUM5 and partly by the H2020 project OPENQKD under grant agreement No. 857156. This work was also supported by the Ministry of Civil Affairs of Bosnia and Herzegovina, under Grant No. 02-7143/20.

\bibliographystyle{unsrt}  
\bibliography{references}  






\end{document}